\documentclass[preprint,showpacs,showkeys,preprintnumbers,amsmath,amssymb]{revtex4}
\usepackage{dcolumn}% Align table columns on decimal point
\usepackage{bm}% bold math
\usepackage{graphicx}
\usepackage{subfigure}
\usepackage{color}

\begin{document}

 \title{The kinetics and its turnover of Hawking-Page phase transition under the black hole evaporation}

\author{Ran Li$^{a,b}$}
\thanks{liran@htu.edu.cn}
\author{Kun Zhang$^c$}
\thanks{zhangkun@ciac.ac.cn}
\author{Jin Wang$^{b,d}$}
\thanks{Corresponding author, jin.wang.1@stonybrook.edu}

\affiliation{$^a$ School of Physics, Henan Normal University, Xinxiang 453007, China}

\affiliation{$^b$ Department of Chemistry, State University of New York at Stony Brook, Stony Brook, NY 11794-3400, USA}

\affiliation{$^c$ State Key Laboratory of Electroanalytical Chemistry, Changchun Institute of Applied Chemistry, Chinese Academy of Sciences, Changchun, China, 130022}

\affiliation{$^d$ Department of Physics and Astronomy, State University of New York at Stony Brook, Stony Brook, NY 11794-3400, USA}

\begin{abstract}
The thermodynamics and the kinetics of Hawking-Page phase transition were studied previously based on the free energy landscape. The AdS black hole can evaporate if imposing the absorbing boundary conditions at infinity. We suggest that the kinetics of Hawking-Page phase transition should be governed by the reaction-diffusion equation, where the Hawking evaporation plays the role of the reaction on the background of the free energy landscape. By calculating the mean first passage time from the large black hole phase to the thermal gas phase, we show that the phase transition can occur more easily under the Hawking radiation. In particular, a kinetic turnover is observed when increasing the ensemble temperature or the frictions. This kinetic turnover can be viewed as the dynamical phase transition to identify the time scale where Hawking evaporation process is comparable to Hawking-Page phase transition.
\end{abstract}

\maketitle

\section{Introduction}

Hawking-Page phase transition \cite{HawkingPage} is the first phase transition between the thermal AdS space and the AdS black hole. In the context of the AdS/CFT correspondence \cite{Maldacena,GKP,Witten}, it was interpreted as the confinement–deconfinement phase transition of the strongly coupled field theory \cite{Wittenphase}. It has been extensively studied in the literature \cite{CKW,Gursoy,EKY,Zhang,Myung,ARS,CaiHP,WSL,Wei:2020kra,Aharony:2019vgs,Mbarek:2018bau,Zhao:2020nrx}. Recently, a first order phase transition in deformed JT gravity, analogous to the Hawking-Page phase transition, was also studied in \cite{Witten:2020ert}.       

However, the dynamics of the Hawking-Page phase transition is still an interesting problem but much less investigated. It is generally believed that above the critical temperature, the large AdS black hole is the dominant phase and the thermal gas in AdS space will collapse to the AdS black hole. Recently, it has been proposed in \cite{Li:2020khm} that the thermodynamics and the kinetics of the black hole phase transition can be studied in terms of the free energy landscape \cite{FSW,FW,NG,JW}. The kinetic process of the phase transition under the thermal fluctuations which is a stochastic process can be properly described by the stochastic Langevin equation or equivalently by the corresponding Fokker-Planck equation \cite{Li:2021vdp}. Based on this proposal, the kinetic time characterized by the mean first passage time can be calculated, which is shown to be closely related to the barrier height on the free energy landscape \cite{Li:2020nsy,Wei:2020rcd,Li:2020spm,Wei:2021bwy}.   

A very important issue ignored in the previous studies \cite{Li:2020khm,Li:2021vdp,Li:2020nsy,Wei:2020rcd,Li:2020spm,Wei:2021bwy} is the Hawking radiation from the black hole horizon \cite{Hawking:1974sw}. In general, the radiation from the AdS black holes will be reflected back into the horizon due to the presence of the timelike boundary. The AdS black holes are in equilibrium states with the thermal radiation \cite{HawkingPage,Witten,Wittenphase}. This is the main reason why the effect of Hawking radiation was not taken into account in the previous works on the kinetics of black hole phase transition. 

In the present work, we consider the absorbing boundary condition or the transparent boundary condition at infinity in AdS \cite{Page:2015rxa,Almheiri:2019psf}. In this case, the Hawking radiation is absorbed by the boundary at infinity or propagates into another universe. As a consequence, no radiation is reflected back to keep the AdS black hole from evaporating. With absorbing or transparent boundary conditions at infinity, the AdS black hole can decay and evaporate through Hawking radiation \cite{Ong:2015fha,Xu:2018liy,Yao:2018ceg,Xu:2019krv,Xu:2020xsl,Hou:2020yni,Wu:2021zyl}. The presence of the Hawking evaporation may significantly change the kinetics of the black hole phase transition. For ordinary materials, it is generally believed that the kinetic rates for most commonly observed first order phase transitions are greatly enhanced by the presence of impurities or imperfections. It is also shown that the decay rate of the false vacuum of the early universe is enhanced by the presence of the black holes as inhomogeneities \cite{Gregory:2013hja,Gregory:2020cvy,Gregory:2020hia}. 

The goal of this paper is to explore the effect of the Hawking evaporation on the kinetics of the Hawking-Page phase transition. We suggest that the kinetics of the Hawking-Page phase transition should be governed by the reaction-diffusion equation \cite{RDE}, where the Hawking evaporation plays the role of the reaction. Based on this proposal, we calculate the mean first passage time from the large black hole phase to the thermal gas phase. It is shown that the phase transition can occur more easily, i.e. the kinetic rates of the phase transition are enhanced by the Hawking evaporation. We also discuss the distributions of the first passage time for the phase transition kinetics and the corresponding fluctuations. In particular, we observed the kinetic turnover behavior of the Hawking-Page phase transition when increasing the temperature or the friction, which can be viewed as the dynamical phase transition.  

This paper is organized as follows. In Sec.\ref{secII}, we review the basics of the free energy landscape description of the Hawking-Page phase transition. In Sec.\ref{secIII}, we consider the Hawking evaporation of the AdS black holes by imposing the absorption boundary at infinity. In Sec.\ref{secIV}, we present the basic assumptions used to study the kinetics of Hawking-Page phase transition when taking the effect of Hawking evaporation into account. In Sec.\ref{secV}, we present the method how to compute the kinetic time of the phase transition characterized by the mean first passage time. In Sec.\ref{secVI}, the numerical results of the phase transition kinetics are presented. The conclusion and discussion are summarized in the last section.  

\section{Free energy landscape description of the Hawking-Page phase transition}\label{secII}

In this section, we review the basic facts of the free energy landscape description of the Hawking-Page phase transition \cite{Li:2020khm}. We are interested in the four dimensional Einstein gravity in AdS space. The formalism can be easily generalized to the modified gravity theories. The metric of the four dimensional Schwarzschild AdS black hole is given by ($G_4 = 1$ units) \cite{HawkingPage}
\begin{eqnarray}
ds^2=-f(r)dt^2+\frac{dr^2}{f(r)}
+r^2\left(d\theta^2+\sin^2\theta d\varphi^2\right)\;,
\end{eqnarray}
where the metric function $f(r)$ is given by
\begin{eqnarray}
f(r)=\left(1-\frac{2M}{r}+\frac{r^2}{L^2}\right)\;.
\end{eqnarray}
The Schwarzschild AdS black hole is determined by only one parameter $M$, which is the black hole mass. The AdS curvature radius $L$ is given by the cosmological constant $\Lambda$ with the relation $L=\sqrt{\frac{-3}{\Lambda}}$.

The black hole horizon $r_+$ is determined by the root of equation $f(r)=0$, i.e.
\begin{eqnarray}
1-\frac{2M}{r}+\frac{r^2}{L^2}=0\;.
\end{eqnarray}
The black hole mass, Hawking temperature, and Bekenstein-Hawking entropy can be expressed in terms of the black hole radius 
\begin{eqnarray}
M&=&\frac{r_+}{2}\left(1+\frac{r_+^2}{L^2}\right)\;,\label{BHmass}\\
T_{H}&=&\frac{1}{4\pi r_+}\left(1+\frac{3r_+^2}{L^2}\right)\;,\label{BHtem}\\
S&=&\pi r_+^2\;.\label{BHS}
\end{eqnarray}
The Hawking temperature has a minimal value
\begin{eqnarray}
T_{min}=\frac{\sqrt{3}}{2\pi L}\;.
\end{eqnarray}

Now we focus on the free energy landscape description of Hawking-Page phase transition \cite{Li:2020khm,Li:2021vdp,Li:2020nsy,Li:2020spm}. It is well known that the thermal fluctuations can give rise to random deviations of a system from the equilibrium state. Black holes are no exception. We treat the thermal gas phase and the black hole phase in AdS space as the local equilibrium states. The fluctuating black holes with arbitrary horizon radius can be generated during the phase transition process. These fluctuating black holes may not be stable. Therefore, the fluctuating black holes are not necessarily the solutions to the Einstein field equations. The canonical ensemble at the specific temperature $T$ can involve the thermal gas phase and the black hole phase in AdS space as well as the fluctuating black holes \cite{Li:2020khm}. The black holes in the ensemble can be distinguished by the continuous order parameter \cite{WeiLiuPRL}, the horizon radius of the black hole. Especially, the thermal AdS gas phase has no horizon, i.e. it is parameterized by $r_+=0$.  

The on-shell Gibbs free energy of the Schwarzschild AdS black hole can be calculated from the Euclidean gravity action. It can be properly rewritten in terms of the thermodynamic relationship of $G=M-T_H S$. In order to describe the fluctuating black holes, we define the generalized off-shell Gibbs free energy as \cite{York:1986it,Whiting:1988qr}
\begin{eqnarray}
G=M-TS=\frac{r_+}{2}\left(1+\frac{r_+^2}{L^2}\right)-\pi T r_+^2\;,
\end{eqnarray}
where $T$ is the temperature of the ensemble. 

\begin{figure}
  \centering
  \includegraphics[width=8cm]{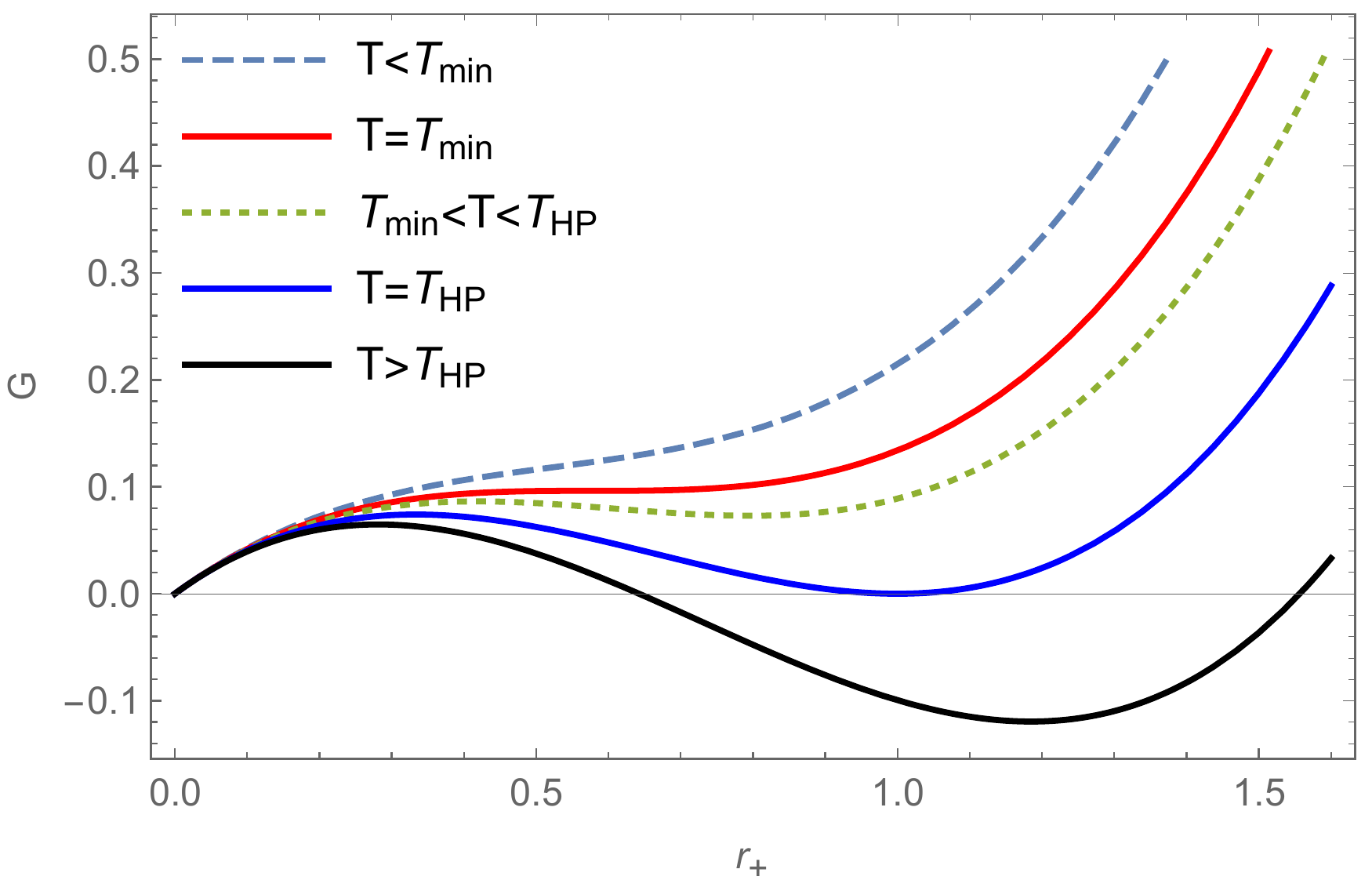}
  \caption{Gibbs free energy landscapes as a function of black hole radius $r_+$ at different temperatures.}
  \label{Gibbs}
\end{figure}

The Gibbs free energy landscapes as a function of the black hole radius $r_+$ \cite{SS,Spallucci} at different temperatures can be plotted, which are explicitly shown in Fig.\ref{Gibbs}. Without the loss of generality, we set the AdS curvature radius $L=1$ in the following. Gibbs free energy landscapes of the present system is only modulated by the temperature. From these plots, we can easily read off the phase diagram of the system as discussed in the following.

When $T<T_{min}$, there is only one global minimum of the Gibbs free energy landscape at the origin, and the system is in a pure radiation phase or thermal AdS space. At $T=T_{min}$, Gibbs free energy landscape exhibits an inflection point at $r=L/\sqrt{3}$. From this temperature on, the two black hole phases emerge (large and small black holes) with radii given by 
\begin{eqnarray}\label{rls}
r_{l,s}=\frac{T}{2\pi T_{min}^2}\left(1\pm\sqrt{1-\frac{T_{min}^2}{T^2}}\right)\;.
\end{eqnarray}
When $T_{min}<T<T_{HP}$, the small black hole phase corresponds to a local maximum of Gibbs free energy, while the large black hole phase is locally stable being a local minimum on the free energy landscape. The globally stable state is still the thermal AdS state. At $T=T_{HP}$, the Gibbs free energy of a black hole phase with radius $r_+=L$ is degenerate with the Gibbs free energy of a thermal AdS phase at the same temperature. $T_{HP}=\frac{1}{\pi L}$ is known as the Hawking-Page critical temperature. Since there is a discontinous change in the order parameter of the thermal AdS phase to that of the large black hole phase on the free energy landscape topography, the associated derivative of the free energy will diverge at the Hawking-Page critical temperature, which is a signature of first order phase transition. Finally, for $T>T_{HP}$ the large black hole phase becomes the absolute minimum of the Gibbs free energy landscape and is globally stable state. In summary, below the critical temperature $T_{HP}$, the thermal AdS phase is thermodynamically stable and above critical temperature the large black hole phase is stable. At the critical temperature, both the thermal AdS space phase and the large black hole phase are stable with equal free energy basin depth.

\section{Hawking evaporation of the Schwarzschild AdS black hole}\label{secIII}

It is known that the AdS black hole is stable due to the timeilke boundary at the spatial infinity. At infinity, one can impose the reflecting boundary to keep the AdS black holes from the evaporation. 
However, Page considered the possibility of imposing the absorption boundary condition to extract the Hawking radiation \cite{Page:2015rxa}. The recent proposal of the transparent boundary conditions and the coupling of AdS black hole to the bath CFT in Minkowski spacetime was essential to the possible resolution of the information paradox \cite{Almheiri:2019psf}. By utilizing the absorption or transparent boundary conditions at the AdS infinity, the Hawking evaporation process of the Schwarzschild AdS black hole can be studied. It is shown by Page that the Schwarzschild AdS black hole has finite lifetime.   

According to  the Stefan-Boltzmann law, the mass of the black hole as a function of time obeys the differential equation
\begin{eqnarray}
\frac{dM}{dt}=-\sigma A T_H^4\;,
\end{eqnarray}
where $\sigma=\frac{\pi^2 k_B^4}{60 c^2 \hbar^3}$ is the Stefan-Boltzmann constant and $A$ is the horizon area of the black hole. Since we are only concerned about the qualitative features of the evaporation process, for simplicity, we use the horizon area as the effective cross section.  

In the next section, we will study the effect of Hawking evaporation on the kinetics of the Hawking-Page phase transition. We treat the Hawking evaporation as an analogue of \textit{chemical reaction}. The reaction rate is defined as the emission rate per unit mass 
\begin{eqnarray}
k=\left|\frac{dM}{Mdt}\right|=\frac{1}{7680\pi} \frac{(1+3r_+^2)^4}{r_+^3(1+r_+^2)}\;,
\end{eqnarray}
where we have set $c=k_B=\hbar=1$. 

In principle, when considering the evaporation effect, the black hole becomes a dynamical object. The energy, the entropy, and the Hawking temperature of the black hole are all time dependent. The free energy landscape description based on the equilibrium assumption for the Hawking-Page phase transition is no longer very accurate. However, noting that the Hawking evaporation process is a relatively slow process compared with the phase transition process and should have a long time scale compared with the kinetic time of the phase transition, we can approximately treat the energy,  the entropy, and the Hawking temperature as quasi-steady in time. Under this approximation, the free energy landscape can still be used to provide a reasonable description. We will check the consistency of this approximate condition according to our numerical results. 

\section{Reaction-Diffusion equation for the stochastic dynamics of Hawking-Page phase transition under Hawking radiation }\label{secIV}

In this section, we will study the kinetics of Hawking-Page phase transition by treating the black hole phase as well as thermal AdS phase as states in a thermodynamic ensemble. We consider there are a large number of states in a thermodynamic ensemble in which one or a group of them can represent Schwarzschild black hole phase or thermal AdS phase or any intermediate transient states during Hawking-Page phase transition. The probability distribution of these states evolving in time should be a function of the order parameter $r_+$ (black hole radius) and time $t$. From now on, we use the symbol $r$ to denote the black hole radius $r_+$ for the sake of simplicity. So the probability distribution is denoted by $\rho(r,t)$. Under the thermal fluctuation and the effect of the Hawking evaporation, the evolution of the probability distribution should be deterministic, although the underlying kinetics is stochastic.  

We propose that the stochastic kinetics of states under the thermal fluctuation and the effect of Hawking evaporation can be described by the reaction-diffusion equation, which on the free energy landscape is explicitly given by
\begin{eqnarray}\label{FPeq}
\frac{\partial \rho(r,t)}{\partial t}=D \frac{\partial}{\partial r}\left\{
e^{-\beta G(r)}\frac{\partial}{\partial r}\left[e^{\beta G(r)}\rho(r,t)\right]
\right\}-k(r)\delta(r-r_l) \rho(r,t)\;.
\end{eqnarray}
In the above equation, the diffusion coefficient $D$ is given by $D=T/\zeta$ with $k$ being the Boltzman constant and $\zeta$ being dissipation coefficient. The inverse temperature of the system is given by $\beta=1/k_BT$. Note that $G(r)$ is the off-shell Gibbs free energy as a function of the black hole radius $r$ modulated by the temperature. If ignoring the last term of the right hand side, the above reaction-diffusion equation can be reduced to the Fokker-Planck equation that describes the kinetics of the black hole phase transition considering only the effect of the thermal fluctuation proposed in \cite{Li:2020khm}. The last term represents the effect of the Hawking evaporation. We call it the reaction term. It describes a decay or sink source for the evaporation process. Because the evaporation process is relatively long, it is expected the reaction term will be significant  at the late stage of the evaporation and comparable to the kinetic time of the phase transition process. The $\delta$-function in the reaction term represents that the evaporation is only effective when the system is in the large black hole state.   

In order to solve the reaction-diffusion equation, two types of boundary conditions should be imposed at the boundaries of the computational domain depending on the question we consider. Note that the reflecting boundary is always imposed at $r=+\infty$, where the free energy is divergent. Therefore, the reaction term will not influence the probability current and the reflecting boundary will not be affected by the reaction term.  

\begin{itemize}
\item Reflecting boundary condition at $r=+\infty$:
\begin{eqnarray}\label{bc1}
\left.
e^{-\beta G(r)}\frac{\partial}{\partial r}\left[e^{\beta G(r)}\rho(r,t)\right]\right|_{r=+\infty}=0\;.
\end{eqnarray}
It is equivalent to
\begin{eqnarray}\label{bc11}
\left.\beta G'(r)\rho(r, t)+\rho'(r, t)\right|_{r=+\infty}=0\;.
\end{eqnarray}

\item Absorbing boundary condition at the small black hole site $r=r_s$:
\begin{eqnarray}\label{bc2}
\rho(r_s,t)=0\;.
\end{eqnarray}
\end{itemize}
With these boundary conditions, we can numerically solve the reaction-diffusion equation to get the time dependent solutions of the probability distribution in the canonical ensemble. Then, we can calculate the kinetic time of the Hawking-Page phase transition, which is characterized by the mean first passage time.

\section{The kinetic time and its fluctuation of the Hawking-Page phase transition}\label{secV}

In general, the kinetic time is characterized by the mean first passage time, which is defined as the mean value of the first passage time of the initial state escaping to the final state \cite{NSM,WangPRE,WangJCP,BW}. In this work, let us consider the first passage time of a state starting from the large black hole phase to the thermal AdS phase. Because it is generally believed that the first passage time of this process can be approximated by the first passage time for a state staring from the large black hole phase to the unstable small black hole state (free energy barrier top from the large black hole phase to the thermal AdS phase). Therefor, we consider the first passage process of a state staring from the large black hole phase to the small black hole state.  

Define $\Sigma(t)$ to be the probability that the state has not made a first passage by time $t$. Suppose there is a perfect absorber placed at the small black hole site $r_s$ where Gibbs free energy attains the local maximum. If the state makes the first passage under the thermal fluctuation, this state leaves the system. Because of the existence of absorber, the normalization of probability distribution will not be preserved in this setup. According to the definition, $\Sigma(t)$ is also the probability of a state being in the system at time $t$. So we have
\begin{eqnarray}\label{Sigmaequation}
\Sigma(t)=\int_{r_s}^{+\infty} \rho(r, t) dr\;.
\end{eqnarray}
At very late time, the total probability of a state still in the system becomes zero, i.e. $\Sigma(r, t)|_{t\rightarrow +\infty}=0$.

As claimed, the first passage time is a random variable. We denote the distribution of first passage times by $F_p(t)$. Then the distributions $F_p(t)$ can be given by
\begin{eqnarray}\label{FPTequation}
F_p(t)=-\frac{d\Sigma(t)}{dt}\;.
\end{eqnarray}
It is obvious that $F_p(t)dt$ is the probability that a state passes through the intermediate small black hole phase for the first time in the time interval $(t, t+dt)$. By substituting Eq.(\ref{Sigmaequation}) into eq.(\ref{FPTequation}), and using the reaction-diffusion equation (\ref{FPeq}), one can get
\begin{eqnarray}\label{FPTcal}
F_p(t)&=&-\frac{d}{dt}\int_{r_s}^{\infty} \rho(r, t) dr\nonumber\\
&=&-\int_{r_s}^{\infty}\frac{\partial}{\partial t} \rho(r, t) dr\nonumber\\
&=&-\int_{r_s}^{\infty}\left\{D\frac{\partial}{\partial r}\left\{
e^{-\beta G(r)}\frac{\partial}{\partial r}\left[e^{\beta G(r)}\rho(r,t)\right]\right\}-k(r)\delta(r-r_l) \rho(r,t)\right\}  dr\nonumber\\
&=&\left.-D e^{-\beta G(r)}\frac{\partial}{\partial r}\left[e^{\beta G(r)}\rho(r,t)\right]\right|_{r_s}^{\infty}+k(r_l)\rho(r_l,t)\nonumber\\
&=&\left.D\frac{\partial}{\partial r}\rho(r,t)\right|_{r=r_s}+k(r_l)\rho(r_l,t)\;.
\end{eqnarray}
Note that the reflecting boundary condition is imposed at $r=+\infty$ because the probability will not leak out there, and the absorbing boundary condition is imposed at $r=r_s$ (transition state at free energy barrier top) where the perfect absorber is placed. By solving the reaction-diffusion equation with the initial condition and the boundary conditions, we can get the time distributions of first passage times. 

In numeric, the $\delta$-function in the reaction-diffusion equation is approximated by the Gauss distribution
\begin{eqnarray}
\delta(r-r_l)=\frac{1}{\sqrt{\pi}b} e^{-(r-r_l)^2/b^2}\;.
\end{eqnarray}

We also choose the initial condition as the Gauss distribution centered at the large black hole site 
\begin{eqnarray}\label{initial}
\rho(r,0)=\frac{1}{\sqrt{\pi}a} e^{-(r-r_l)^2/a^2}\;.
\end{eqnarray}

With the time distributions, we can calculate the mean first passage time and its fluctuation. The mean first passage time is defined by
\begin{eqnarray}\label{MFPTcal}
\langle t \rangle=\int_{0}^{+\infty} t F_p(t) dt\;.
\end{eqnarray}
In principle, we can also calculate the $n$-th moment of time distribution function of first passage time by the relation
\begin{eqnarray}\label{nthmoment}
\langle t^n \rangle=\int_{0}^{+\infty} t^n F_P(t) dt\;.
\end{eqnarray}
In consequence, we can calculate the fluctuation of the kinetic time which is given by $\left(\langle t^2 \rangle-\langle t \rangle^2\right)$. 

\section{Numerical Results}\label{secVI}

\begin{figure}
  \centering
  \includegraphics[width=8cm]{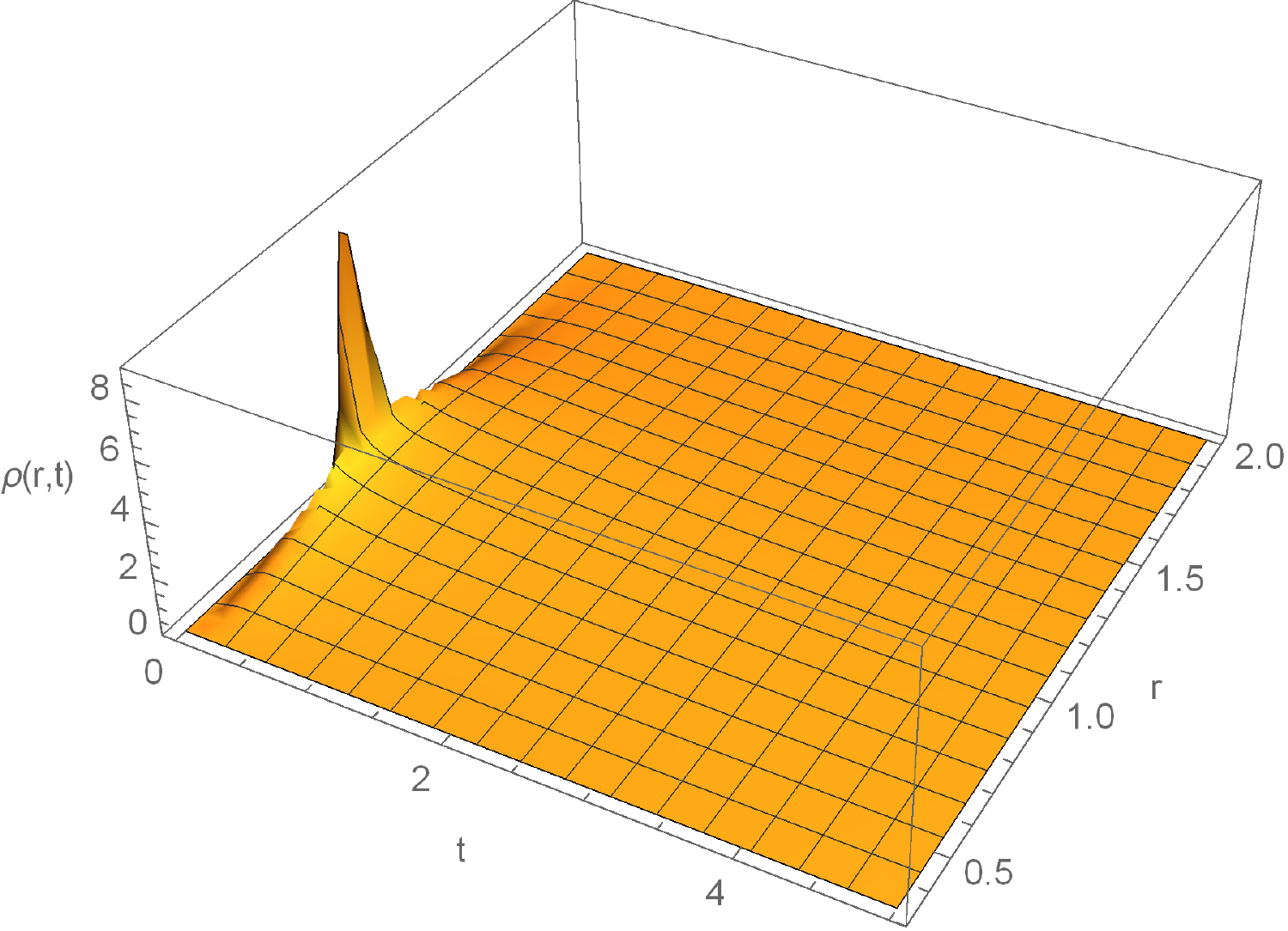}
   \includegraphics[width=8cm]{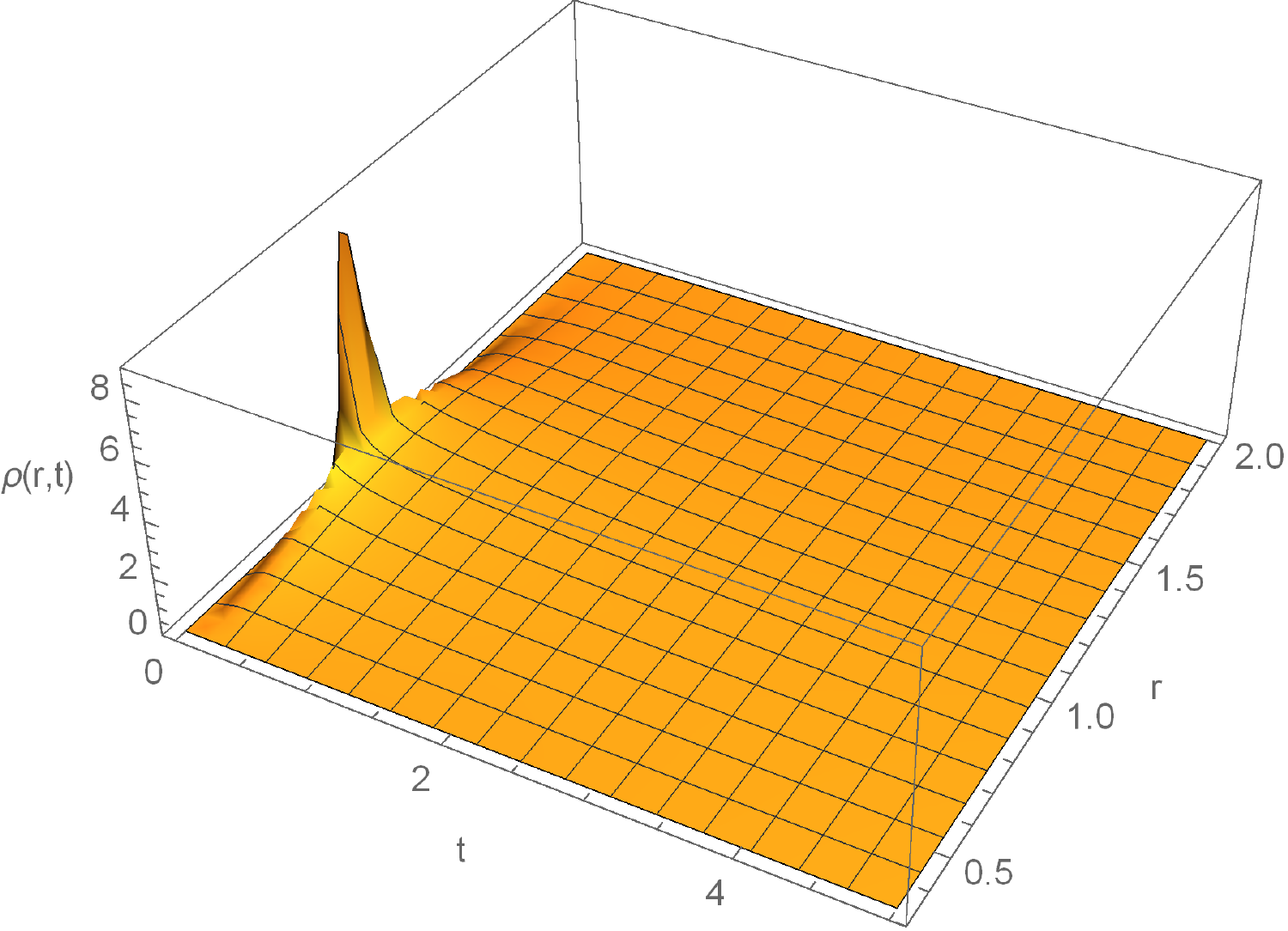}
  \caption{The time evolution of the probability density $\rho(r,t)$ at the phase transition point $T=T_{HP}$. In the plots, $\zeta=1$. The left panel is the plot without the evaporation reaction and the right for the evaporation reaction. }
  \label{rhoplot}
\end{figure}
In Fig.\ref{rhoplot}, the time evolution profiles of the probability distribution $\rho(r,t)$ without (left panel) and with the reaction term are plotted. The temperatures are taken to be the Hawking-Page critical temperature. Due to the absorption boundary condition at the top of the free energy barrier, the probability density becomes zero at the late time. In fact, because of the low barrier height between the large black hole state and the small black hole state, the probability density becomes to zero in a very short time. By comparing the probability evolution with and without the evaporation reaction, we can observe that the reaction term does not influence the probability distribution significantly at the phase transition point. This is consistent with our previous assumption that the Hawking evaporation process is a relative slow process compared with the phase transition process. This point can be further verified by the time distributions of the phase transition kinetics with and without the evaporation reaction. 

\begin{figure}
  \centering
  \includegraphics[width=8cm]{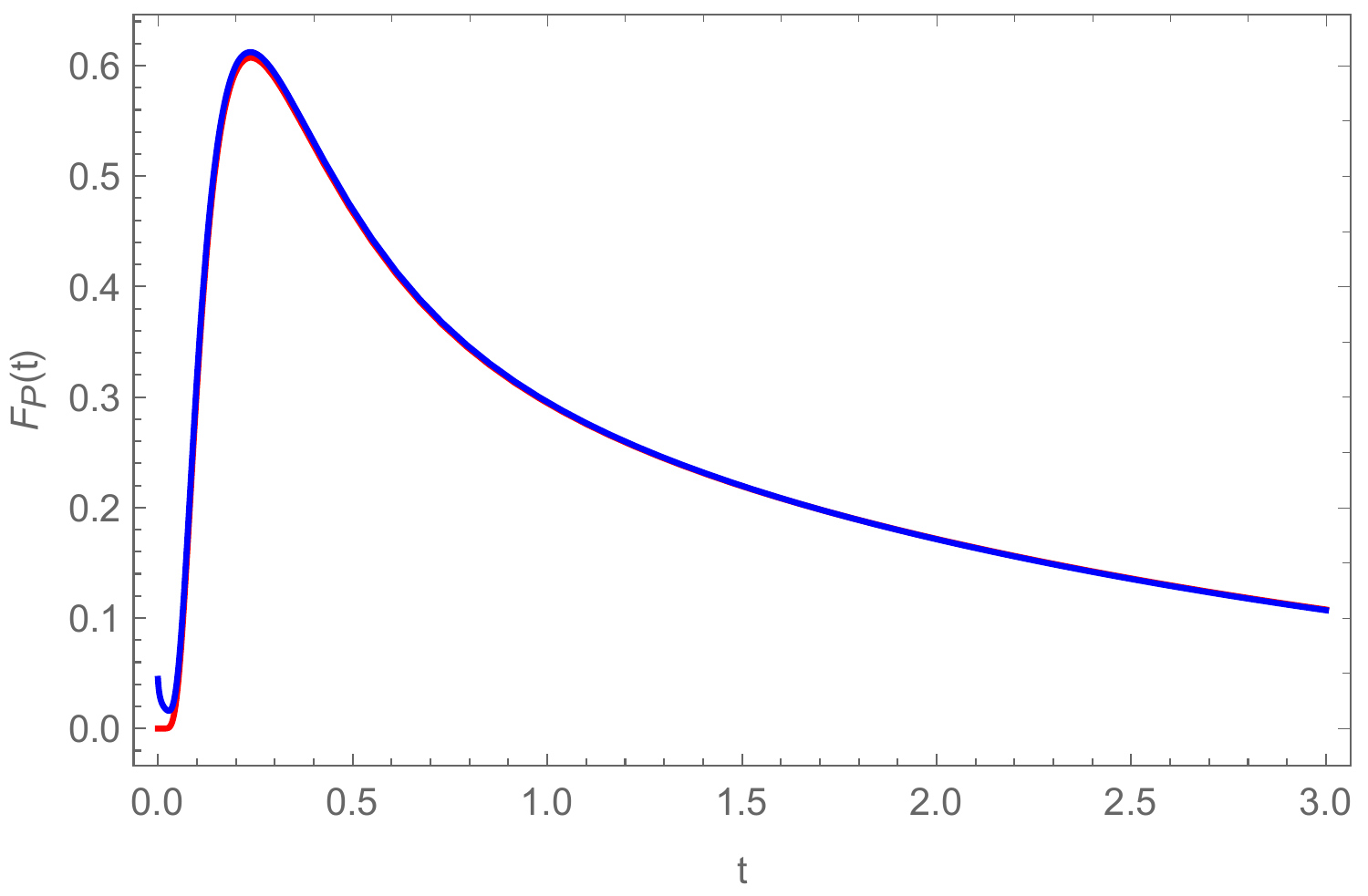}
    \caption{The time distributions $F_p(t)$ without the evaporation reaction (red curve) and with the evaporation reaction (blue curve) at the Hawking-Page critical temperature.}
  \label{FptPlotTHP}
\end{figure}

In Fig.\ref{FptPlotTHP}, we have plotted the distributions of the first passage time corresponding to the cases in Fig.\ref{rhoplot}. Without the reaction term, $F_p(t)$ starts with zero at $t=0$, goes to a maximum value at a finite time, and decays to zero at late time. With the evaporation reaction, the numerical result shows that $F_p(t)$ is not zero at $t=0$. This is caused by the nonzero reaction term at $t=0$ in Eq.(\ref{FPTcal}). The time distribution with the reaction term (blue line) is not influenced by the reaction term for large $t$. The barrier height between the small black hole and the large black hole is very low, which results the small mean first passage time for the kinetics. The reaction term does not seem to influence the kinetic process of the Hawking-Page phase transition. This implies that as long as the mean first passage time of the phase transition process is short enough, the effect of the evaporation reaction is rather limited.    

\begin{figure}
  \centering
  \includegraphics[width=8cm]{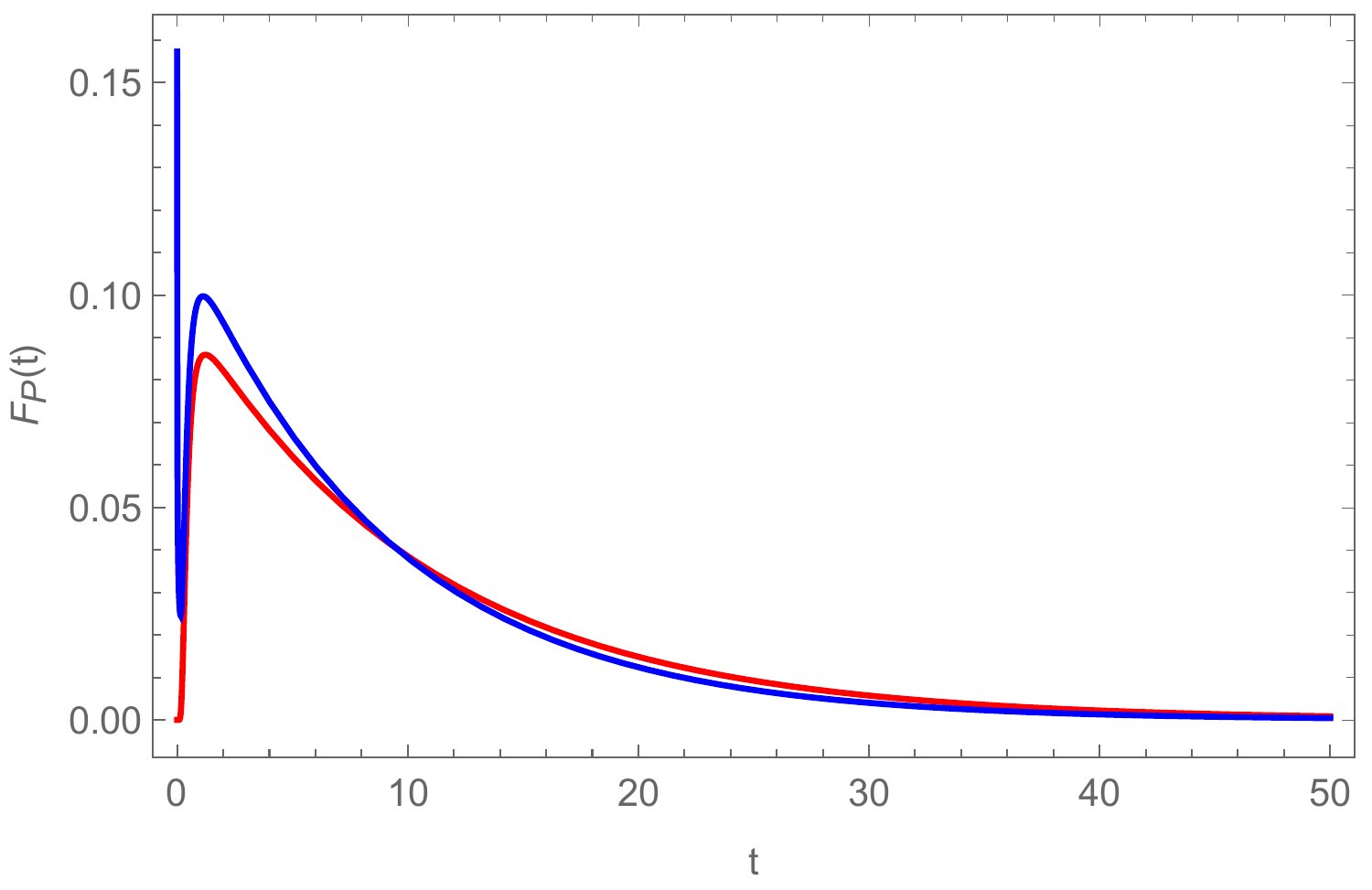}
  \caption{The time distribution $F_p(t)$ without the evaporation reaction (red curve) and with the evaporation reaction (blue curve) at the temperature $T=0.45$. }
  \label{FptPlotT045}
\end{figure}
In Fig.\ref{FptPlotT045}, we plot the time distributions at the temperature $T=0.45$, which is greater than the Hawking-Page critical temperature. In this case, the kinetic time is longer than that of the previous case, because the barrier height becomes larger. The results indicate that the time distribution is significantly influenced by the reaction term when the temperature becomes higher beyond the Hawking-Page critical temperature. This also implies that the evaporation reaction can significantly change the kinetics of the phase transition when the barrier height between the large black hole state and the thermal AdS phase is relatively high (when the temperature becomes higher beyond Hawking-Page transition temperature).   

\begin{figure}
  \centering
  \includegraphics[width=8cm]{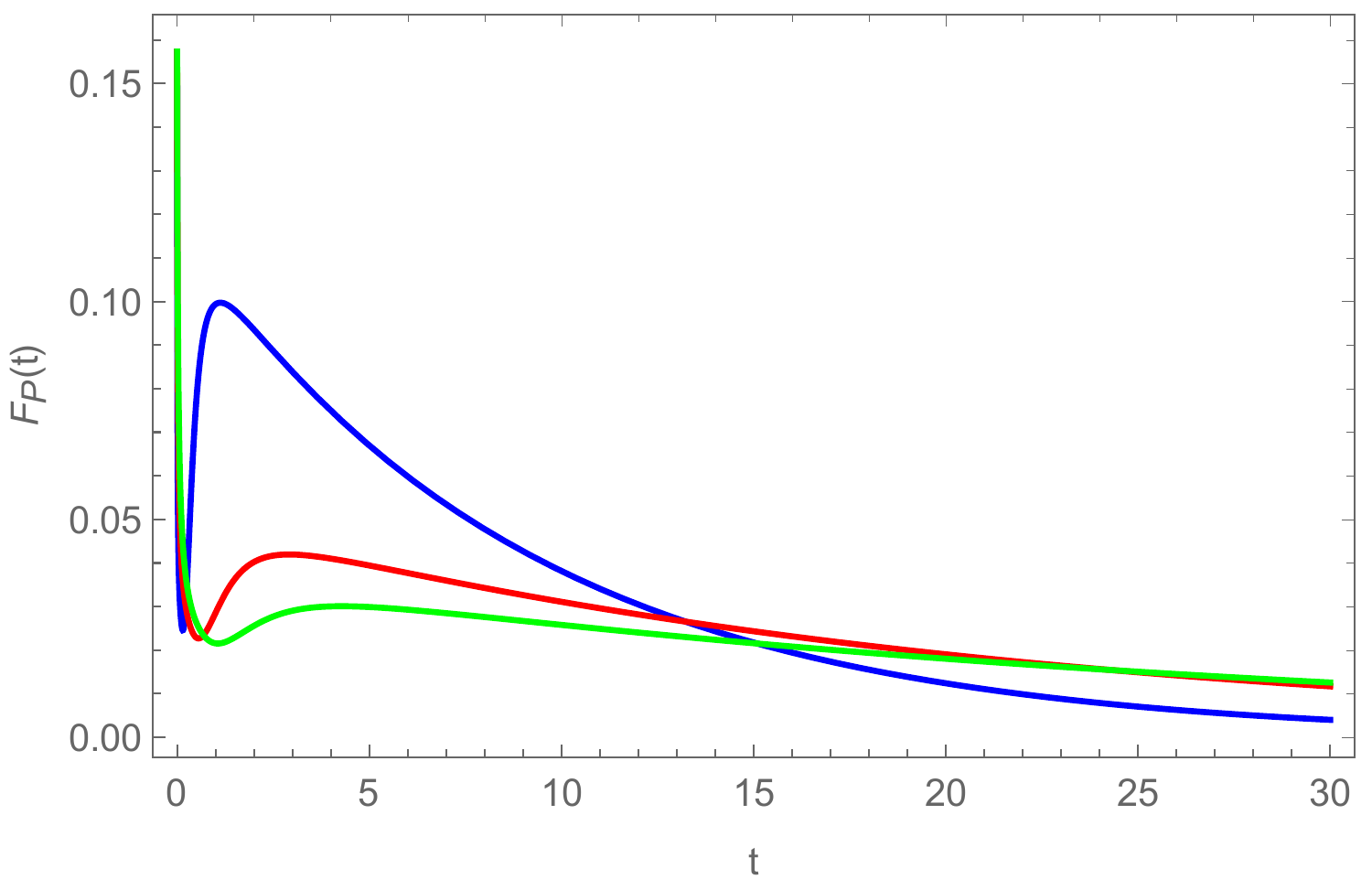}
  \caption{The time distribution $F_p(t)$ with the evaporation reaction for different friction coefficients. $T=0.45$ and the red for $\zeta=1$, the blue for $\zeta=3$, and the green for $\zeta=5$.}
  \label{FptPlotdiffzeta}
\end{figure}
In Fig.\ref{FptPlotdiffzeta}, we have present the time distributions for different friction coefficients at the temperature $T=0.45$. When $\zeta$ increases, the time distribution becomes wider. Therefore, the mean first passage time increases correspondingly.  Without the evaporation reaction, when increasing the friction coefficient, the time distributions also becomes wider. The reason behind is that the kinetic time is proportional to the friction coefficient in the over damped regime. We can also observe that the effect of the evaporation reaction becomes more significant when increasing the friction coefficient. Therefore, we can see that the evaporation process is often a relatively slow process at this temperature compared to the phase transition process at the normal friction. However, when the friction coefficient becomes larger, the evaporation process can be relevant to the phase transition process.  

\begin{figure}
  \centering
  \includegraphics[width=8cm]{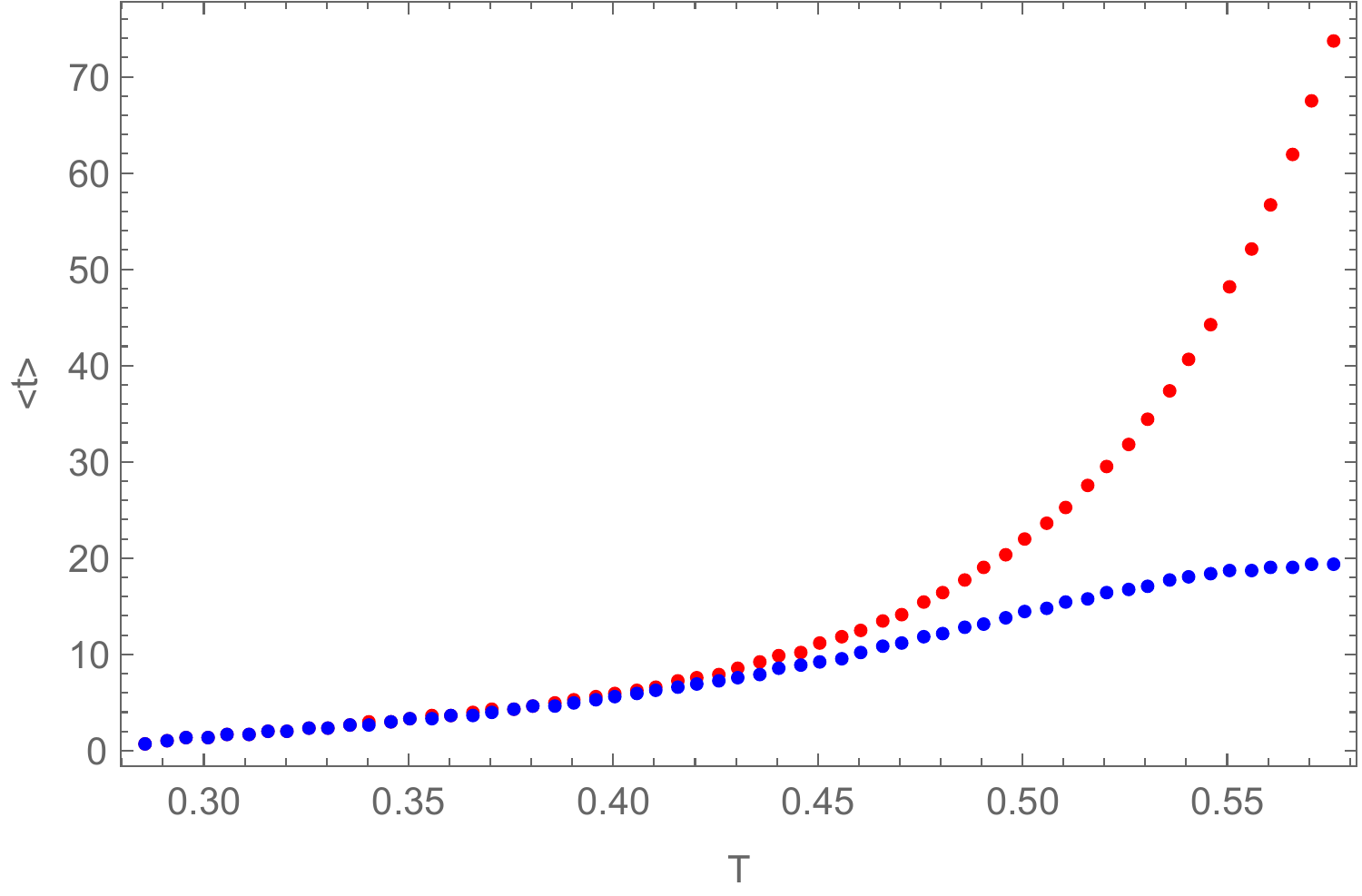}
   \includegraphics[width=8cm]{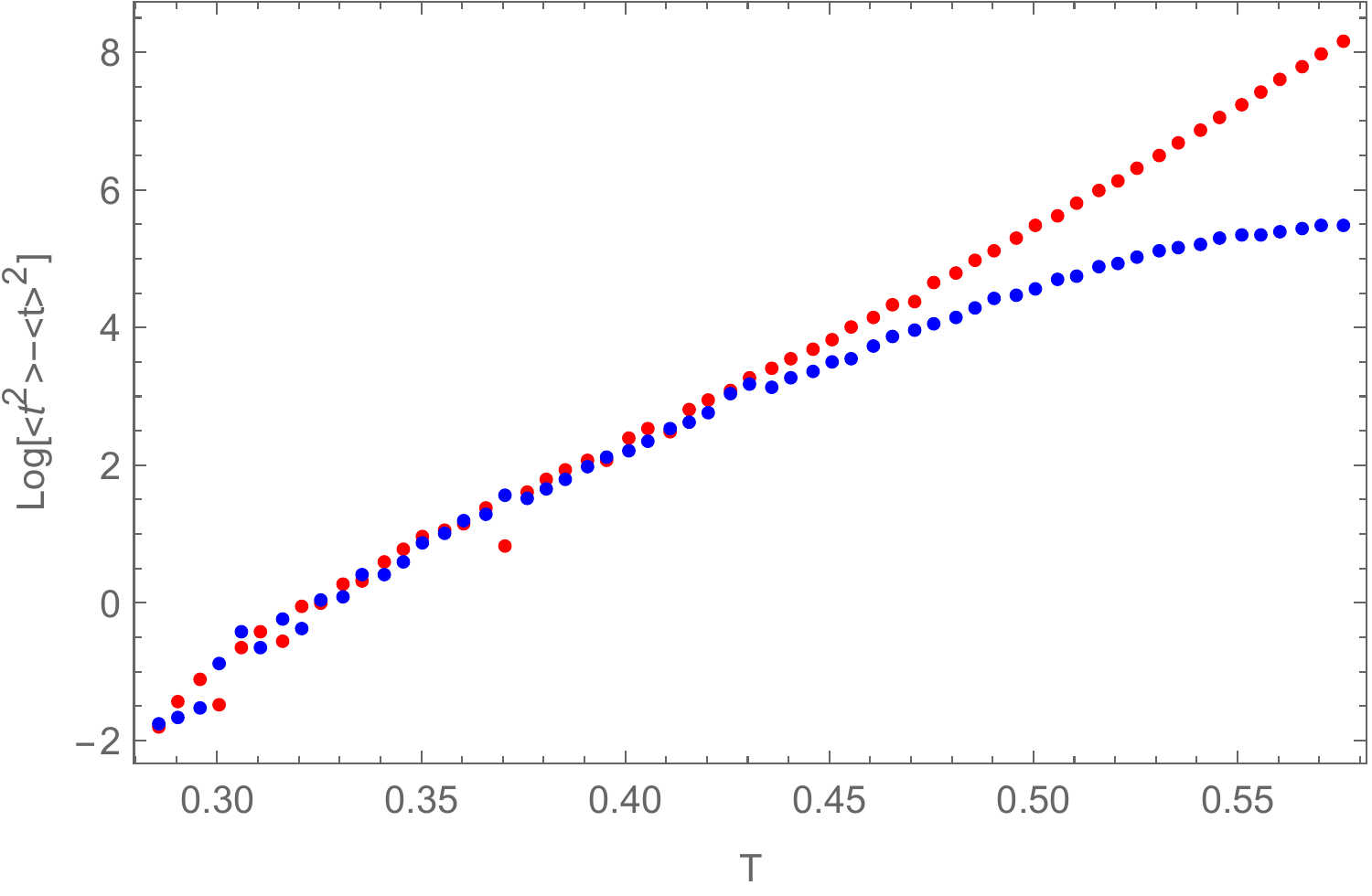}
  \caption{The MFPT and the fluctuation as the functions of temperature with the friction coefficient fixed. In the plots, $\zeta=1$. The red dotted curve is the result calculated from the Fokker-Planck equation without the evaporation reaction, while the blue dotted curve is the result from the reaction-diffusion equation with the evaporation reaction. }
  \label{MFPTvsT}
\end{figure}
In the left panel of Fig.\ref{MFPTvsT}, the MFPTs as the function of temperature for the Fokker-Planck equation without the reaction and for the reaction-diffusion equation are plotted.  When the barrier height is low, the MFPT is small. The effect of the evaporation reaction is not significant. When increasing the temperature, the barrier height between the small black hole and the large black hole increases. This leads to the long MFPT. It can be seen that the reaction has significant influence on the kinetics of the phase transition process. Because the reaction can accelerate the process of the probability density $\rho(r,t)$ approaching zero, the MFPT for the kinetics with the evaporation reaction is smaller than that without the reaction. In the right panel Fig.\ref{MFPTvsT}, the corresponding fluctuations are presented. The fluctuation is a monotonic increasing function of the temperature. In other words, the Hawking radiation process can accelerate the black hole phase transition from the large black hole to the thermal AdS phase.  Since the corresponding kinetic fluctuation under the Hawking radiation is smaller, this indicates that the Hawking evaporation can facilitate the black hole phase transition robustly.

\begin{figure}
  \centering
  \includegraphics[width=8cm]{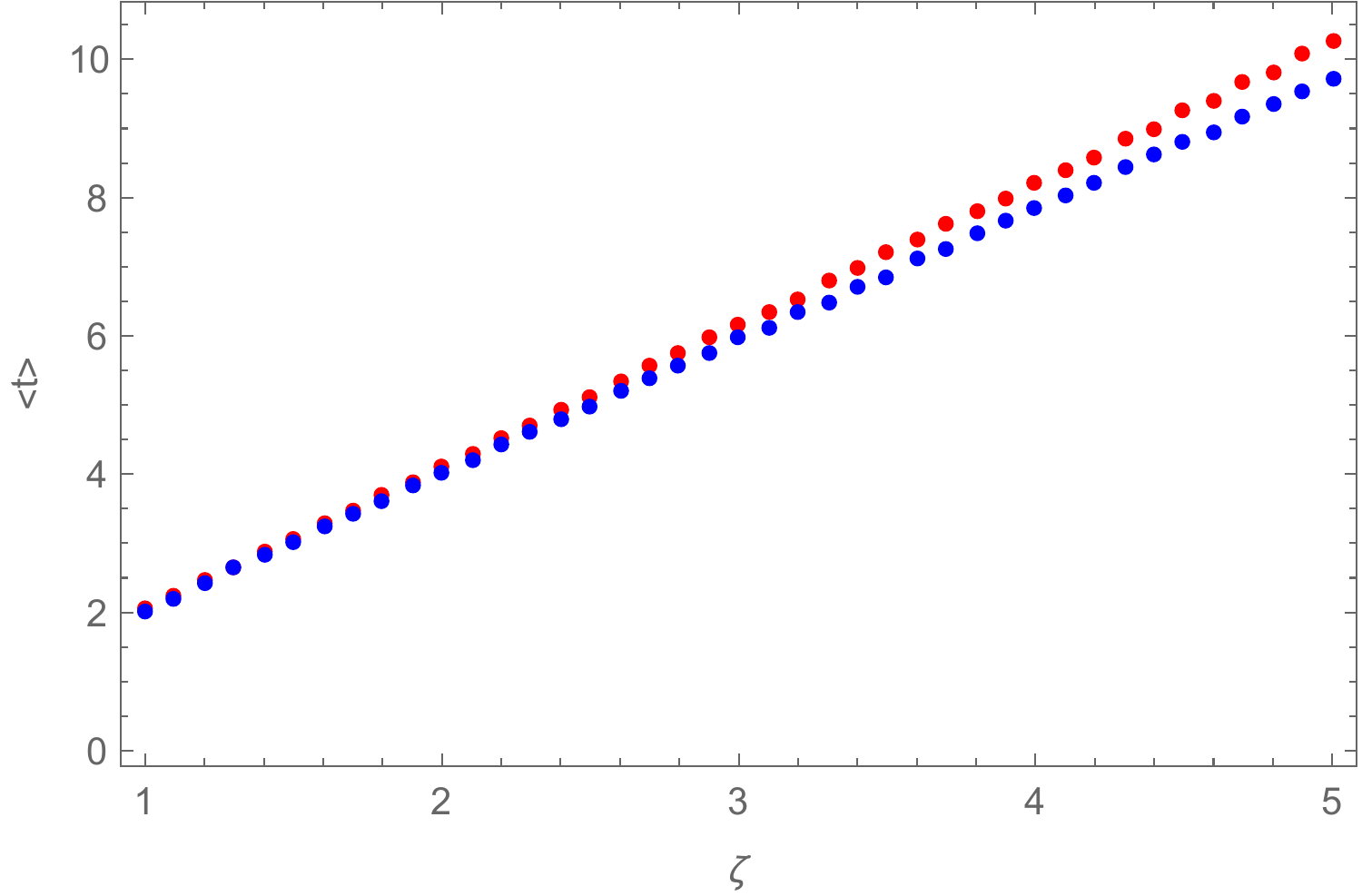}
   \includegraphics[width=8cm]{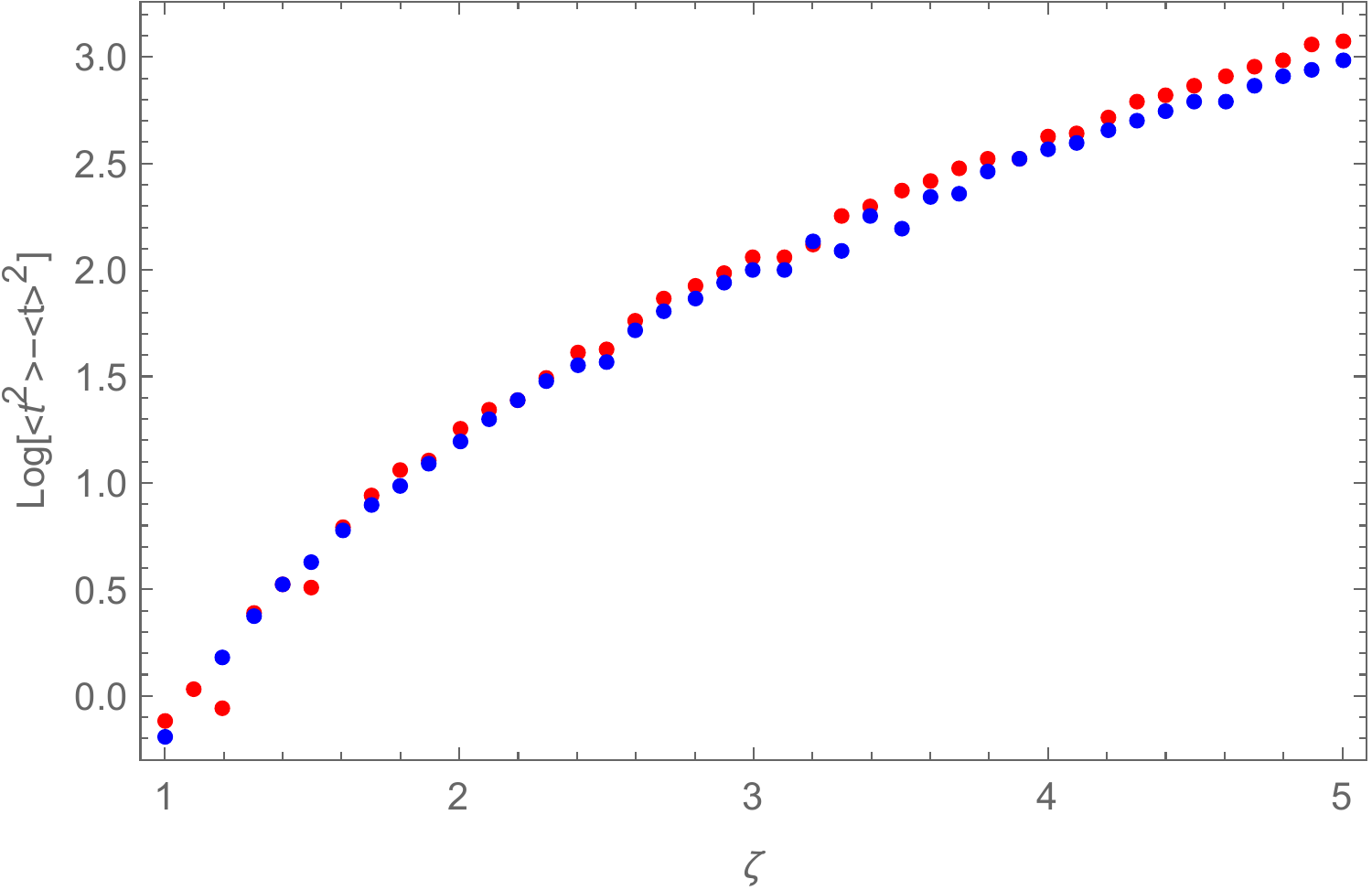}
  \caption{The MFPT and the fluctuation as the functions of the friction coefficient with the temperature fixed. In the plots, $T=T_{HP}$. The red dotted curve is the result calculated from the Fokker-Planck equation without the evaporation reaction, while the blue dotted curve is the result from the reaction-diffusion equation with the evaporation reaction.  }
  \label{MFPTvsZetaTHP}
\end{figure}

In Fig.\ref{MFPTvsZetaTHP}, the MFPT and the fluctuation as the functions of the friction coefficient with the  fixed temperature $T=T_{HP}$ are plotted. In this case, the effect of the evaporation reaction is not significant. The numerical results show that the MFPT is proportional to the friction coefficient. This is a sign of the system in the over-damped regime.

\begin{figure}
  \centering
  \includegraphics[width=8cm]{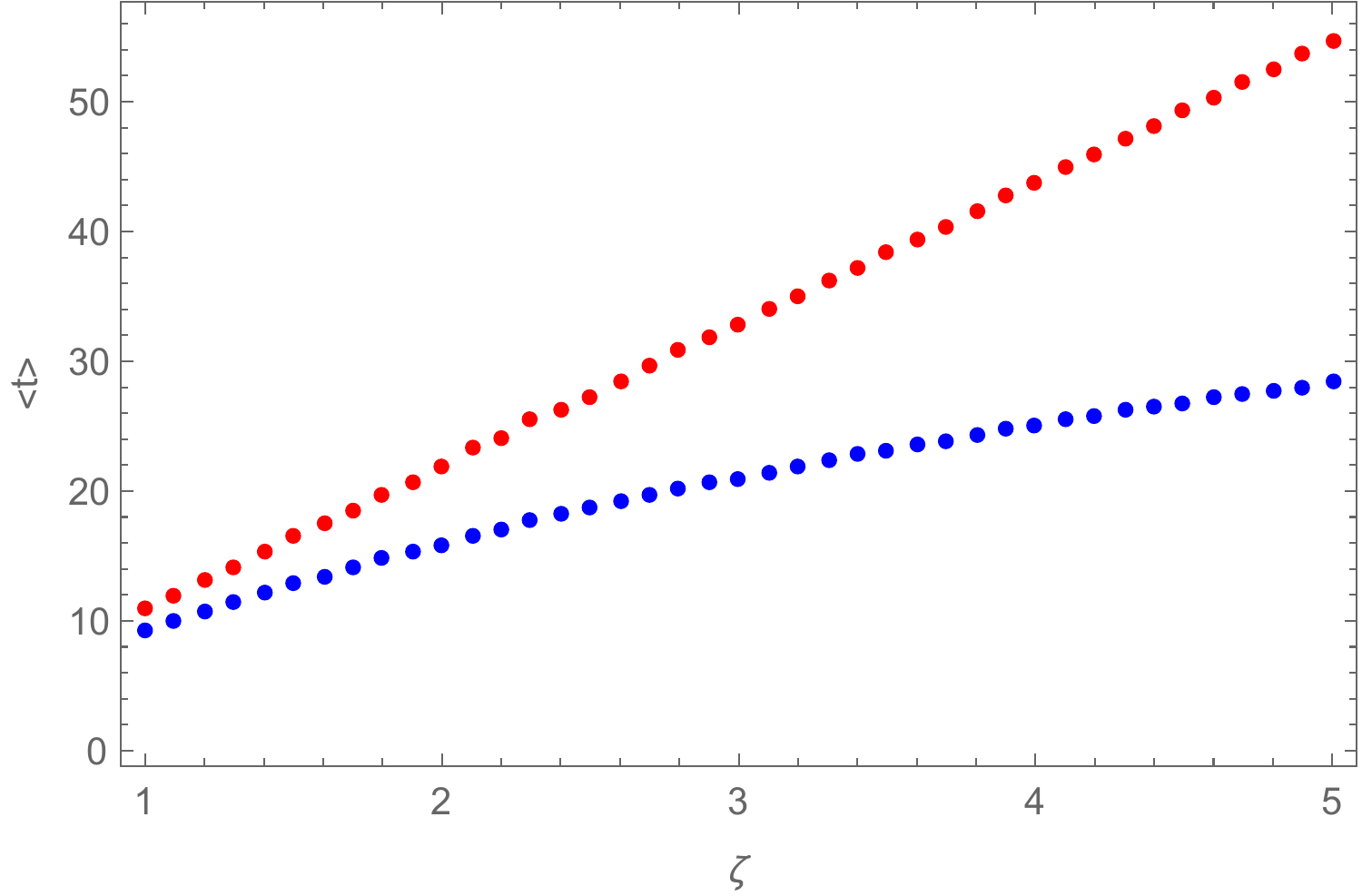}
   \includegraphics[width=8cm]{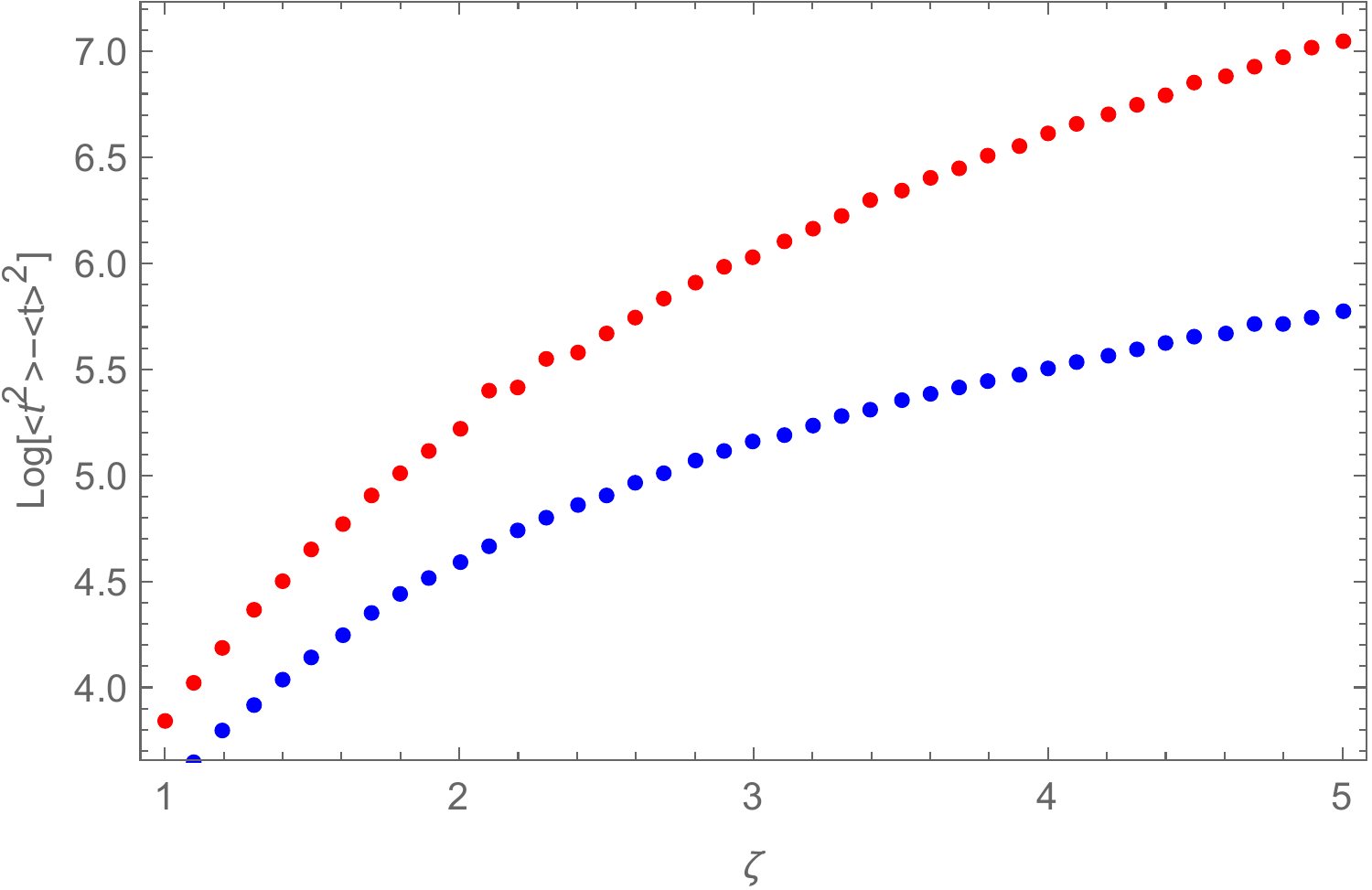}
  \caption{The MFPT and the fluctuation as the functions of the friction coefficient with the temperature fixed. In the plots, $T=0.45$. The red dotted curve is the result calculated from the Fokker-Planck equation without the evaporation reaction, while the blue dotted curve is the result from the reaction-diffusion equation with the evaporation reaction. }
  \label{MFPTvsZetaT045}
\end{figure}
In Fig.\ref{MFPTvsZetaT045}, the MFPT and the fluctuation as the functions of the friction coefficient under the  fixed temperature $T=0.45$ are plotted. In this case, the effect of the evaporation reaction is very significant. Without the reaction term, the MFPT is proportional to the friction coefficient. While for the reaction-diffusion process, the kinetics is influenced by the evaporation reaction significantly. The resulting kinetics of the black hole phase transition from the large black hole to the AdS thermal phase is faster with less fluctuation than the case when the Hawking evaporation is not taken into account.

\begin{figure}
  \centering
  \includegraphics[width=8cm]{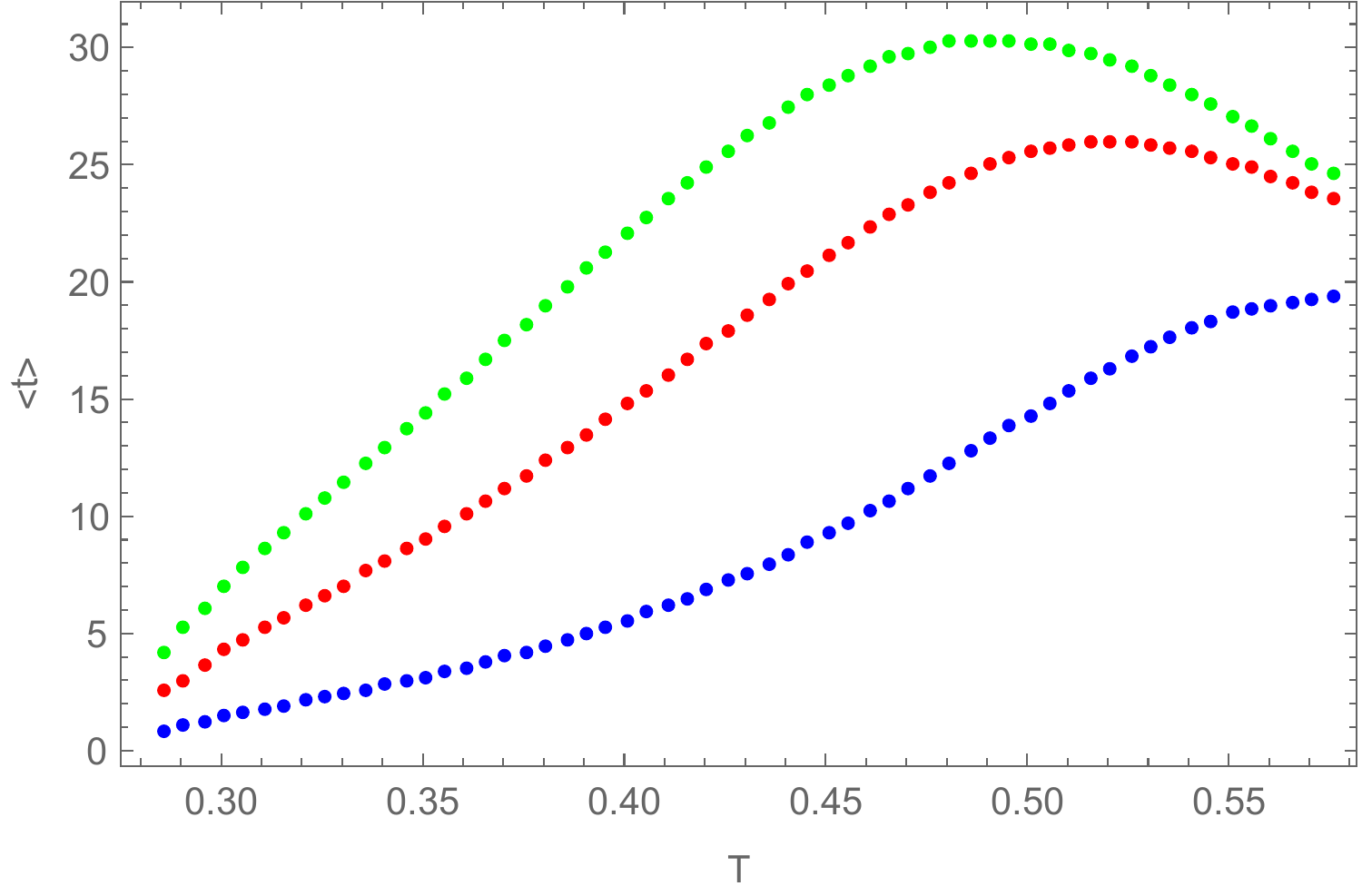}
   \includegraphics[width=8cm]{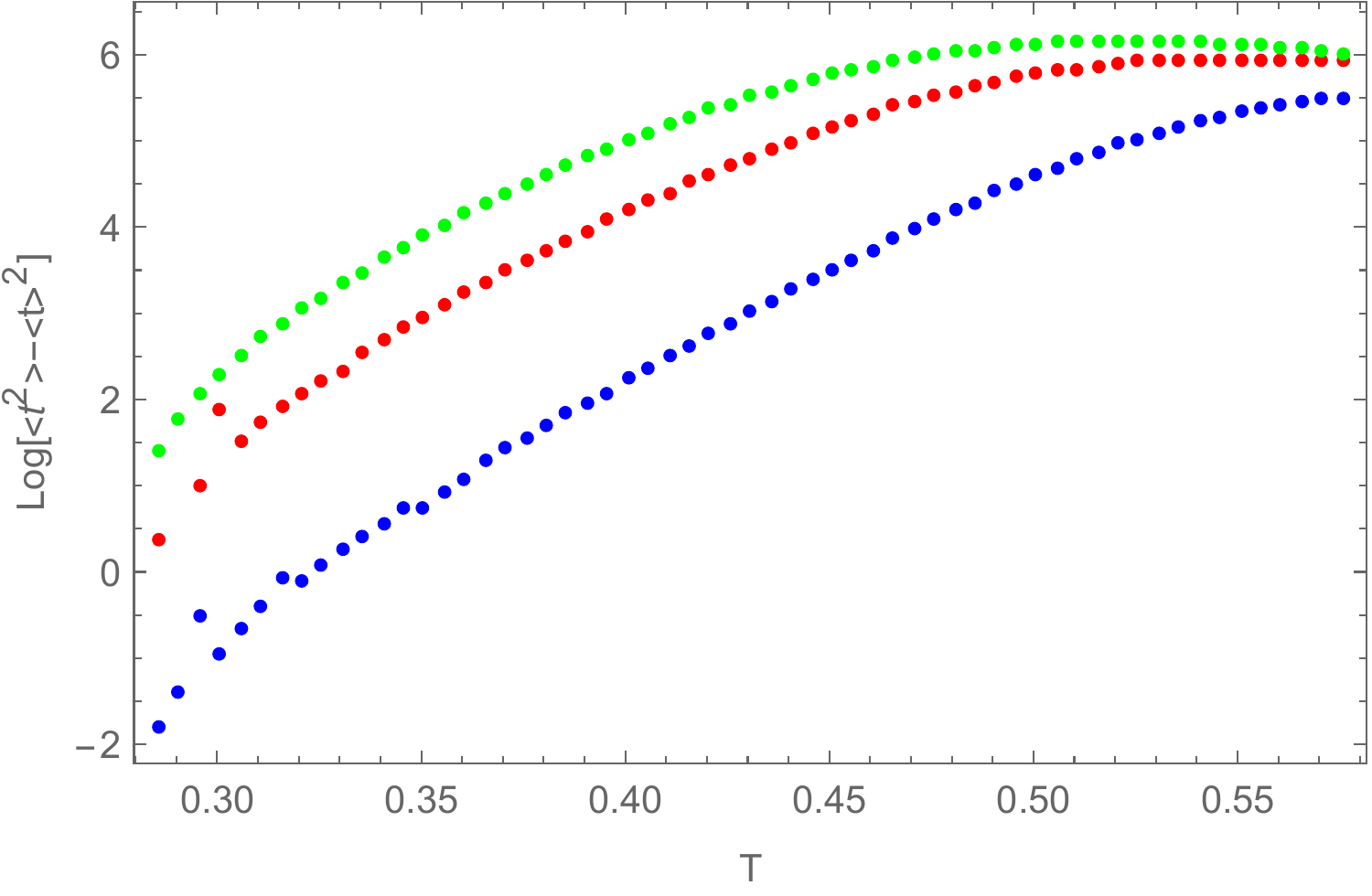}
  \caption{The MFPT and the fluctuation as the functions of the temperature at different friction coefficients. $\zeta=1 (blue), 3 (red), 5 (green)$. The results are calculated by using the reaction-difusion equation. }
  \label{MFPTvsTdiffzeta}
\end{figure}

In Fig.\ref{MFPTvsTdiffzeta}, the MFPT and the fluctuation as the functions of the temperature at different friction coefficients are plotted. For larger friction coefficient, the effect of the evaporation reaction on the kinetics is more significant. For the large friction coefficients, we have observed the kinetic turnover ($T\sim 0.5-0.58$ for blue, red, and green curves respectively). In fact, in the whole parameter range of the friction coefficient, we have observed the turnover phenomenon. This is very different from the conclusion in \cite{Li:2021vdp}, where a kinetic turnover is observed when increasing the friction coefficient. According to the previous results, increasing the temperature can lead to the higher barrier, and in consequence a long mean first passage time. However, this conclusion appears to be violated when taking the evaporation reaction into account. In other words, when increasing the temperature, the barrier from the large black hole to the thermal AdS space becomes higher and the corresponding switching time becomes longer. As the switching time scale is comparable to the evaporation reaction time scale, the reaction can effectively influence the switching or phase transition dynamics. Specifically, the evaporation reaction can accelerate the phase transition process. Therefore, when increasing the temperature, the kinetic time is the result of the competition between the barrier height and the evaporation reaction. Beyond certain temperature, evaporation effect dominates and leads to faster phase transition dynamics. This temperature is the kinetic turnover temperature.

\begin{figure}
  \centering
  \includegraphics[width=8cm]{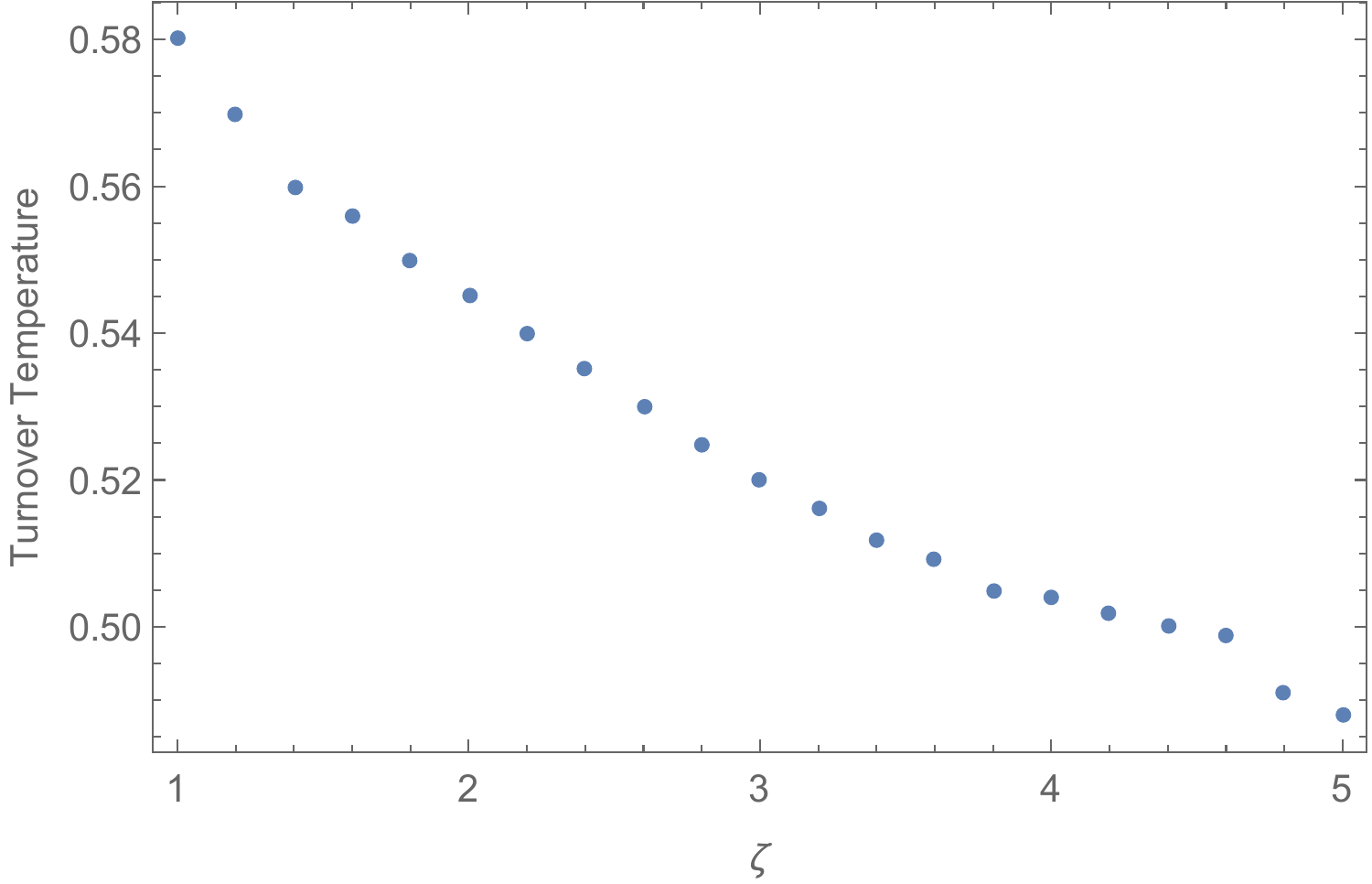}
  \caption{Phase diagram of the kinetic turnover temperature versus the friction coefficient. }
  \label{TurnTemp}
\end{figure}

The plot of the turnover temperature versus the friction coefficient is presented in Fig.\ref{TurnTemp}. This provides an effective phase diagram. The dotted line separate the $T-\zeta$ plane into two parts or two kinetic phases. In the left lower part of the dynamical phase diagram, the Hawking-Page phase transition process dominates. We have the usual kinetic behavior that the higher barrier slows the kinetics. In the right upper part of the dynamical phase diagram, the Hawking evaporation process dominates. The Hawking evaporation is comparable or faster compared with the phase transition process and becomes more significant. The effect of the evaporation is to reduce the chance of staying in the large black hole state. This effectively facilities the transition from the large black hole to the thermal AdS phase. The evaporation dominated phase leads to the kinetics of the phase transition which is not dictated by the barrier height of the underlying free energy landscape. Therefore, we observe a dynamical phase transition from the free energy landscape determined kinetics to the evaporation dominated kinetics in certain temperature and friction ranges.

\begin{figure}
  \centering
  \includegraphics[width=8cm]{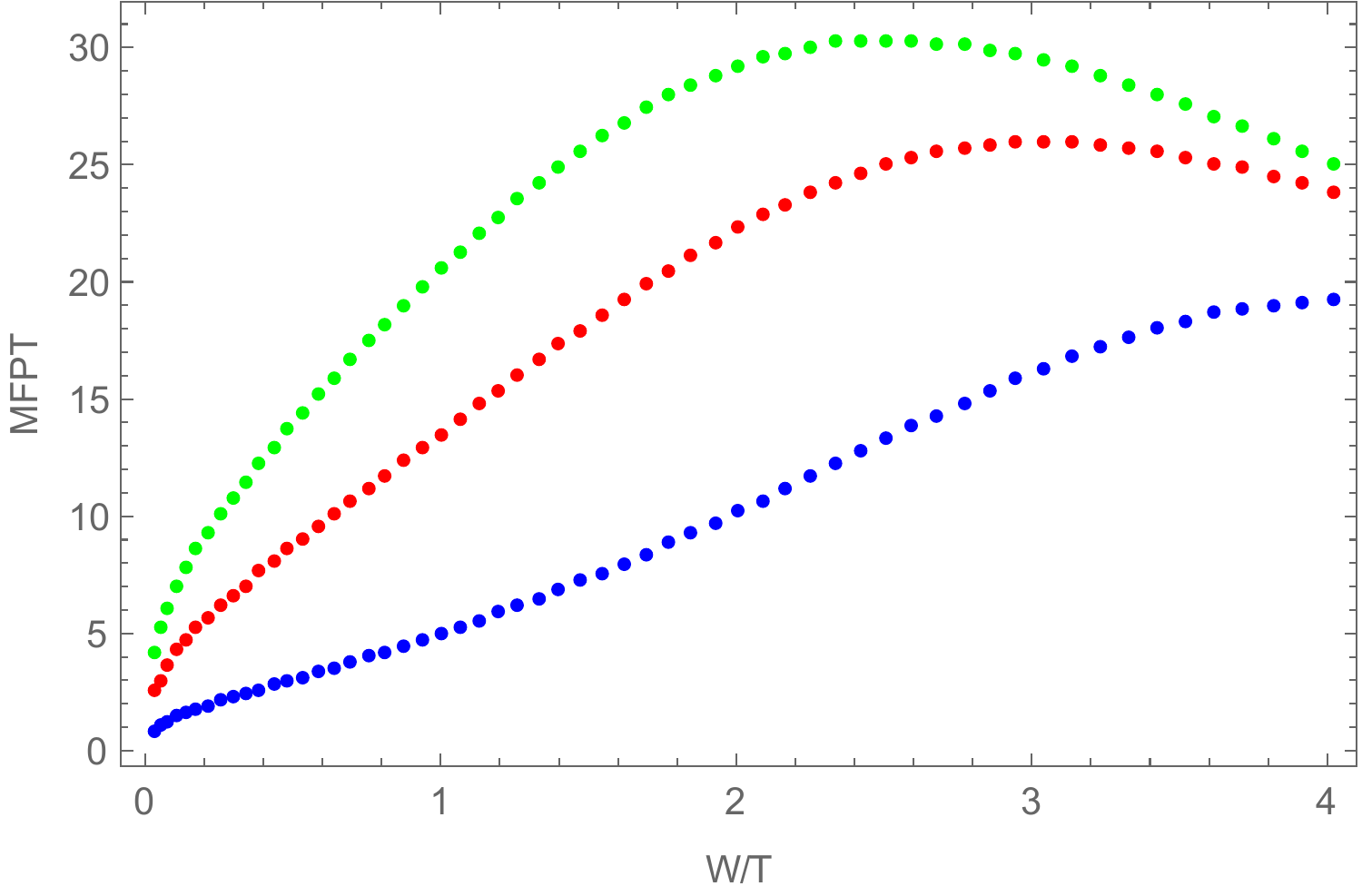}
  \caption{The MFPT versus the ratio of the barrier height and temperature for $\zeta=1 (blue), 3 (red), 5 (green)$. }
  \label{MFPTvsBarrier_Height}
\end{figure}

However, we should point out that the phase diagram is not exact, because the reaction rate $k$ is taken as a constant. This is a reasonable approximation when the Hawking-Page phase transition process dominates. When the evaporation process dominates, the large AdS black hole is a dynamical object. The reaction rate $k$ is also a time dependent function of the radius of the large AdS black hole. This implies in the right upper part of the dynamical phase diagram as shown in Fig.\ref{TurnTemp}, the kinetic time or the MFPT should get significant correction if considering the time dependence of the reaction rate. For the left lower part of the dynamical phase diagram, we argue the behavior of the MFPT is determined by the barrier height mostly, that is $W=G(r_s)-G(r_l)$. In order to show the correlation of the kinetic time and the barrier height, in Fig.\ref{MFPTvsBarrier_Height}, we have plotted the MFPT versus $W/T$ for different friction coefficients. It is shown that before the kinetic turnover point ($W/T\sim2-4$ for blue, red, and green respectively) the MFPT increases monotonically with respect to the barrier height $W/T$. Notice that the analytical calculation has shown that the kinetic time without the evaporation reaction is determined by the barrier height via the relation $\langle t \rangle \sim e^{W/T}$ \cite{Li:2021vdp}. It appears that the Hawking effect (the evaporation reaction) changes the kinetics of the black hole phase transition significantly. After the kinetic turnover point, the MFPT is not an increasing function of the $W/T$ any more. This implies that the evaporation process dominates the phase transition process.

\section{Conclusion and discussion}\label{secVII}

In summary, we have studied the Hawking evaporation effect on the kinetics of the Hawking-Page phase transition. We suggest that the kinetics of Hawking-Page phase transition should be governed by the reaction-diffusion equation, where the Hawking evaporation plays the role of the reaction on the background of the free energy landscape of the black hole phase transition. Based on this proposal, we calculate the mean first passage time from the large black hole phase to the thermal gas phase. It is shown that the phase transition can occur more easily under the Hawking radiation. Especially, a kinetic turnover is observed when increasing the ensemble temperature. The kinetic turnover point can be used to identify the time scale where Hawking radiation is comparable to phase transition especially at late times of evaporation. This kinetic turnover can be viewed as the dynamical phase transition for the black hole from the kinetics dictated by the free energy landscape barrier to the kinetics dominated by the evaporation process. 

We believe that studying the kinetics of Hawking-Page phase transition, in conjunction with the previously discussions on the black hole phase transition based on the free energy landscape \cite{Li:2020khm,Li:2021vdp,Li:2020nsy,Wei:2020rcd,Li:2020spm,Wei:2021bwy}, can lead to progresses on the probing and understanding black hole microstructures. This work provides a two dimensional probe to the black hole microstructure: changing friction to probe the interaction strengths of the black hole microstructures and changing the temperature to probe the kinetic turnover from the Hawking radiation reflecting the black hole microscopic degrees of freedom. These studies can be generalized to investigate the Hawking-Page phase transitions or black hole phase transitions in the extended phase space or in the modified gravity theories.

 \end{document}